\documentclass[apj]{emulateapj}  
\usepackage{graphics} 
\usepackage{apjfonts}
\usepackage{mathptmx}
\usepackage{psfig}
\newcommand {\lsim}{\mbox{$\:\stackrel{<}{_{\sim}}\:$} }
\newcommand {\gsim}{\mbox{$\:\stackrel{>}{_{\sim}}\:$} }
\newcommand {\etal}{{et\thinspace al.} }
\newcommand {\parn}{\par\noindent}
\newcommand{\F}{\mbox{$\cal F$}}

\def\Om{\Omega_m}
\def\Ob{\Omega_{\rm b}}
\def\Ol{\mbox{$\Omega_{\Lambda}$}}

\def\iso#1#2{\mbox{${}^{#2}{\rm #1}$}}
\def\li#1{\hbox{$^{#1}{\rm Li}$}}
\def\be#1{\iso{Be}{#1}}
\def\b#1#2{\iso{B}{#1#2}}
\def\6Li{\mbox{$^6$Li}}
\def\7Li{\mbox{$^7$Li}}
\def\he#1{\iso{He}{#1}}
\def\igm{intergalactic medium}
\def\zs{\mbox{$z_{\rm s}$}}
\def\zgal{\mbox{$z_{\rm gal}$}}

\def\zstar{\mbox{$z^\star$}}
\def\Eps{\mbox{$E_s^{'}$}}

\def \Sec#1{{Sec\-tion~\ref{s:#1}}}
\def \Tab#1{{Table~\ref{t:#1}}}
\def \Eq#1{{Eq.~\ref{e:#1}}}     
\def \EQN#1{\label{e:#1}}        
\def \Fig#1{{Fig.~\ref{f:#1}}}   

\def \unit#1#2{#1$^{#2}$}
\def \d#1{{\rm d} #1}

\shorttitle{}
\shortauthors{}

\begin{document}

\journalinfo{
\parbox{1.5in}{
astro-ph/0412426 \\
UMN--TH--2335/04 \\
FTPI--MINN--04/47 \\
December 2004}
} 

\title{Cosmological Cosmic Rays and the observed $^6$Li plateau in metal poor halo stars}

\author{Emmanuel Rollinde\altaffilmark{1}, Elisabeth Vangioni\altaffilmark{2}, Keith Olive\altaffilmark{3}}
\altaffiltext{1}{IUCAA, Post Bag 4, Ganesh Khind, Pune 411 007, India, rollinde@iucaa.ernet.in}
\altaffiltext{2}{Institut d'Astrophysique de Paris, 98 bis bd Arago, 75014 Paris, France, vangioni@iap.fr}
\altaffiltext{3}{William I. Fine Theoretical Physics Institute, School of Physics and Astronomy, University of Minnesota, Minneapolis, MN 55455, USA, olive@physics.umn.edu}

\begin{abstract}
\parn 

 Very recent observations of the \6Li\ isotope in halo stars
  reveal  a \6Li\  plateau about 1000 times above the 
 predicted BBN abundance. 
 We calculate the evolution of \6Li\ versus
  redshift generated from an initial burst of cosmological 
cosmic rays (CCRs) up to 
the formation of the Galaxy.
We show that the  pregalactic production of the \6Li\
 isotope  can account for the \li6
  plateau observed in metal poor halo stars
without  additional over-production of \li7.
The derived relation between
the amplitude of the CCR energy spectra and the redshift of 
the initial CCR production puts constraints on the physics
and history of the objects, such 
as pop III stars, responsible for these early cosmic rays. 
Consequently, we consider the evolution of \6Li\ in the Galaxy.
Since \6Li\ is also produced in Galactic cosmic ray nucleosynthesis, 
we argue that halo stars with metallicities between [Fe/H] = -2 and -1,
must be somewhat depleted in \6Li.
\end{abstract}


\section{Introduction}
\label{s:introduction}

To account for the origin and evolution of lithium, beryllium and boron, we rely on 
our understanding of several very different aspects of nucleosynthesis, namely:
Big Bang, non thermal,  stellar nucleosynthesis, all of which must be
correlated through cosmic and chemical evolution. 
These rare light nuclei are not generated in the normal course 
of stellar nucleosynthesis (except \7Li\ in the galactic disk) and 
are in fact destroyed in stellar interiors. 
This explains the relatively low abundance of these species. 
While a significant fraction of the observed \7Li\ is produced in the Big Bang, the
Big Bang nucleosynthesis (BBN) production of 
\6Li, Be and B results in abundances which are orders of magnitude
below that observed in halo stars. For example,  BBN production of \6Li\ is 
dominated by the process D$(\alpha,\gamma)$\6Li. At the baryon density
deduced from observations of the anisotropies of the Cosmic Microwave Background
 (CMB) radiation by WMAP \citep{Spergel}, its BBN value is \6Li/H $\simeq$ $10^{-14}$ 
 \citep{tsof,Vangioni99}.
 On the other hand the BBN mean value of the \7Li\ abundance
 is, according
to \citet{Cyburt04} \7Li/H = $4.27^{+1.02}_{-0.83}\times 10^{-10}$, according to 
\citet{Cuoco}   \7Li/H = $4.9^{+1.4}_{-1.2}\times10^{-10}$, or
  according to \citet{Coc} \7Li/H = $4.15^{+0.49}_{-0.45}\times10^{-10}$.
As such,  the \7Li/\6Li\ ratio in BBN is about $4 \times 10^4$.
 
The very low abundances of the \6Li, $^{9}$Be
 and $^{10,11}$B 
isotopes predicted by
BBN theory imply that their most plausible production process   
was the interaction of  Galactic Cosmic rays (GCRs) 
 with the interstellar medium 
\citep[for a review see][]{Vangioni00}. Of these isotopes,  \6Li\ is of particular interest because
 it has only recently been measured in halo stars 
 \citep{sln1,ht1,ht2,sln2,cetal,Nissen99, Nissen00,Asplund1,Asplund2, Aoki} 
 thus offering new constraints on the very early evolution of light elements \citep{sfosw}. 
 Many studies have followed the evolution of \6Li\ in our Galaxy 
 \citep[see e.g.][]{Fields99,Vangioni99}.
 Of particular importance in this context is 
the $\alpha+\alpha$ reaction that 
leads to the synthesis of this isotope (as well as \7Li)
and is efficient very early in the evolutionary history of the Galaxy.

Different scenarios have been
discussed to explain the abundance of \6Li\ in metal-poor
halo stars (MPHS). \citet{Suzuki} discussed  
the possibility of cosmic rays produced in shocks during
the  formation of the Galaxy, which was
consistent with \6Li\ data available at that time.
\citet{Jedamzik} considers
the decay of relic particles, during  the epoch of the big bang nucleosynthesis,
that can yield to a large primordial abundance of \6Li.
\citet{Fields} have  studied in detail the lithium
production in connection to gamma rays, using a formalism similar
to ours but with a different point of view as far as the observational
constraints are concerned
(see \Sec{comparison}).

Until recently, the abundance of \6Li\ had 
been observed in only a few MPHS with metallicity [Fe/H] larger
than -2.3. New values of 
the ratio \6Li/\7Li have been measured with UVES at the 
VLT-UT2 Kueyen ESO telescope, 
 in halo stars with metallicity 
ranging from  -2.7 to -0.5 (see \Sec{observation}).
 These observations indicate the presence of a plateau in \6Li/H $\simeq\, 10^{-11}$
which suggests a pregalactic origin for the formation of \6Li.

In this paper, we consider the synthesis of lithium due to 
the interaction of cosmological cosmic rays (CCRs),
 produced at an early epoch, with the intergalactic medium 
\citep[and \Sec{formalism}]{Montmerlea,Montmerleb,Montmerlec}.
 As $\alpha + \alpha$ processes also produce \li7,
these models are constrained by the \li7 plateau observed in the same MPHS.
This constraint is made more severe by the current discrepancy between the
BBN predicted value of \li7 and the observational abundance.
We demonstrate how this model can explain 
the recent observations 
and constrain the history of cosmological structure formation
(Section~\ref{s:result}) and the galactic evolution of \6Li\ 
(Section~\ref{s:discussion}).
We compare these results with the expected evolution of \li6 from
GCR nucleosynthesis.  Without the pregalactic production of \li6, the latter model
can not account for the elevated \li6 abundances at very low metallicity. In contrast, 
models for which O/Fe increases at low metallicity are able to produce
sufficient \li6 at low metallicity, without pregalactic production.  
However, in this case, the bulk of the
\li6 data seen in higher metallicity stars must be argued to be depleted.
The same is true for our model of CCR nucleosynthesis, but to a lesser extent.
We argue that \li6 data in stars with [Fe/H] = -3 to -4 will
be required to distinguish between these scenarios.
Our predictions will be compared 
to other work in \Sec{comparison} and 
our conclusions are given in \Sec{conclusion}.

\section{Observational and nuclear data}
\label{s:observation}

The determination of the \li6 abundance in MPHS is
extremely difficult and requires high resolution and high signal to
noise spectra due to the tiny hyperfine splitting between the two
lithium isotopes. The 
line splitting is only seen as the narrowly shifted lines are thermally broadened.
Though the fits to the width of this feature are sensitive to the \li7/\li6 ratio,
it is very difficult to obtain accurate measurements of the isotopic ratio.
\li6 can only be realistically expected to be
observed in stars with high surface temperatures and at
low metallicities of [Fe/H] $ <  -1.3$.
\citet{bs} determined that only in stars with surface
temperatures greater than about 6300 K will \li6 survive in the
observable surface layers of the star.  At metallicities [Fe/H] $\ga -1.3$, even
higher effective temperatures would be required to preserve \li6.

\begin{figure}[!h]
\unitlength=1cm
\begin{picture}(12,9)
\put(0,0){\psfig{width=\linewidth,figure=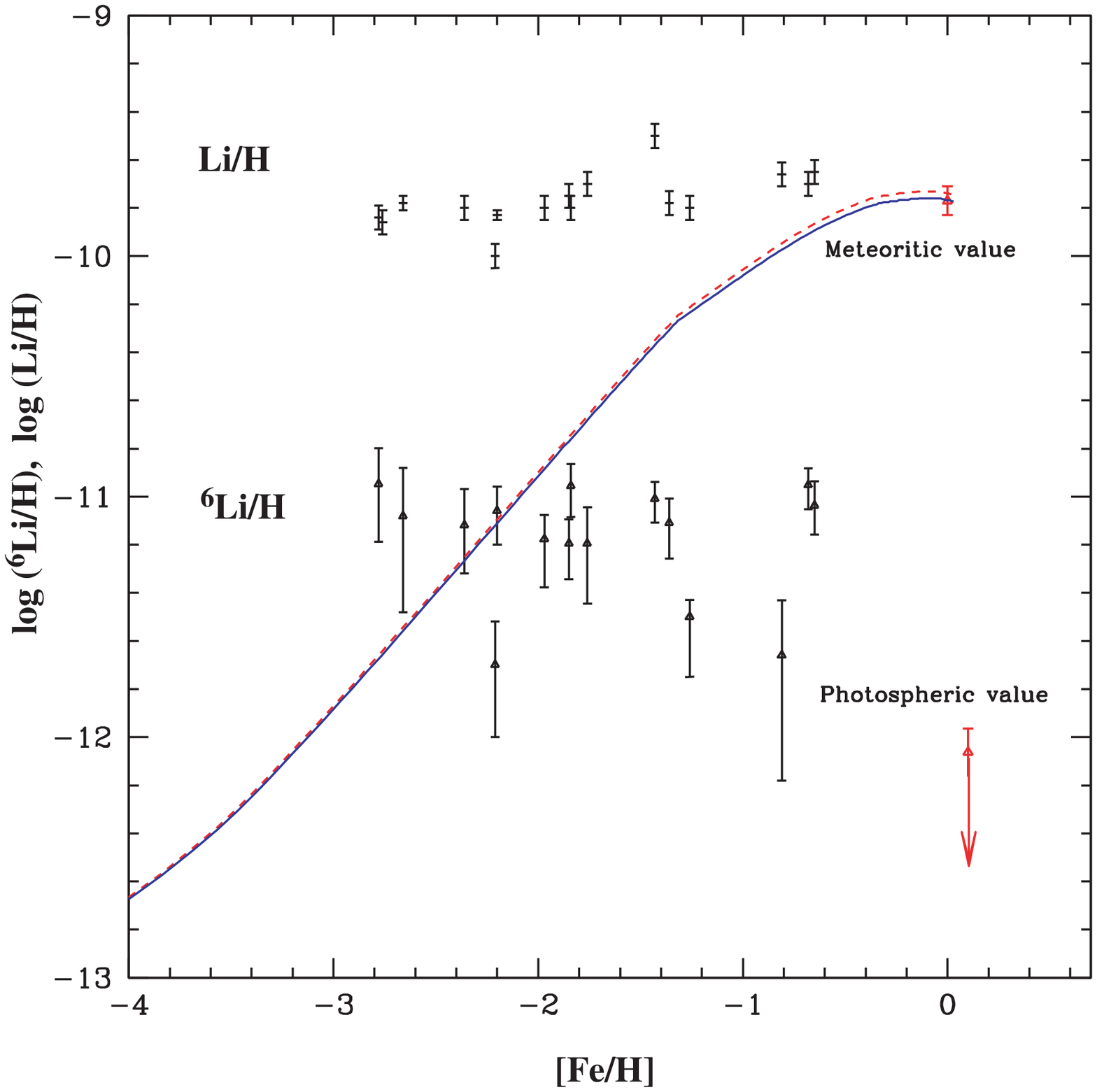}}
\end{picture}
\caption{
The evolution of \6Li/H vs [Fe/H]. 
Here, the evolution of \6Li/H is modeled by GCR nucleosynthesis alone.
 Predictions using  cross sections from \protect\citet{Mercer}
 (lower line) are reduced compared to those from  \protect\cite{read84} 
(upper dashed line). The abundances of \6Li\ 
in low metallicity stars reveal a plateau.
The solar abundance of \li6  from 
meteorites \protect\citep{Lodders03}  and the upper limit to the 
solar photospheric abundance \protect\citep{Asplund3} 
are also shown.
}
\label{f:obs}
\end{figure}

As noted above, the previous sets of data on 
the lithium isotope ratio has been significantly expanded
by \citet{Asplund2} \citep[see also][]{Lambert}
with the observations of 24 MPHS.
Previously, only 3 stars with metallicity [Fe/H] $ < -1.3$
showed net detections of \li6
\citep{sln1,ht1,ht2,sln2,cetal,Nissen00}.
The observed abundances  of Li/H and \6Li/H 
are displayed versus the metallicity, [Fe/H], in \Fig{obs}. 
There are in addition several stars with metallicities in the 
range [Fe/H] = -3 to -0.5 for which only upper limits (not shown) to the
\li6/\li7 ratio are available.
Note that the \6Li\ abundance
at solar metallicity is plotted for both the meteoritic value
\citep{Lodders03} and  the solar 
photospheric value \citep{Asplund3}. The latter is 
derived from the photospheric value of Li assuming the
solar ratio \li7/\li6 =12 and therefore really represents an upper
limit to the \li6 photospheric abundance.

These new data at low metallicity reveal 
the existence of a plateau for \6Li,
whose abundance is about 1000 times higher than that
predicted by BBN. 
As such,  another  production
mechanism which is capable of producing what appears to be 
an initial enrichment of \li6 in the intergalactic medium is required.
Here, we concentrate 
on the interaction of $\alpha$ particles
  present in CCRs produced at high redshift,
with He at rest in the IGM,
as a potential description of this pregalactic enrichment process.
The abundances in
higher metallicity stars will be discussed in \Sec{discussion}.

Note that there is, however, considerable dispersion in the 
data, and as noted above, there are many stars for which 
there was no detectable \li6, indicating that depletion may have 
played a role in the observed \li6 abundance for stars in the plateau as well.
This is in contrast to the \li7 plateau which shows very little dispersion and
for which we expect the role of depletion to have been minor \citep{rbofn}.

Galactic production of lithium
arises from the interaction of GCRs with the ISM via the $\alpha + \alpha$
reaction. This is a primary  process which yields  
a logarithmic slope of 1 in the lithium abundance versus metallicity ([Fe/H]) relation
as seen in \Fig{obs}.
The amplitude for \li6 production is  constrained
by the abundances of Be and B (see \Sec{discussion} for an additional
discussion).  As one can see, this model can not explain the elevated \li6 abundances
at low metallicity and requires some \li6 depletion at higher metallicity (as
will all of the models discussed here). 

Recently, \citet{Mercer} have performed new measurements related to
the $\alpha+\alpha$ reaction and provide a new fit for the production of  \6Li\
and \7Li. 
The calculated \6Li\ abundances from the GCR process 
using the often applied \citet{read84} cross sections
and  the fit at higher energy provided by \citet{Mercer} resulting in slightly
less (30 -- 50 \% at solar metallicity)  \li6 
are compared in \Fig{obs}. 
In what follows, we use the most recent cross sections of \citet{Mercer}.

\section{CCR production of lithium in the IGM : Formalism}
\label{s:formalism}

\subsection{On the Existence of Cosmological Cosmic Rays}

The existence and global properties of  cosmic rays
in the Galaxy 
are often related to supernova explosions and/or gamma-ray bursts,
 in massive stars.
Motivated by the WMAP results indicating an early epoch of reionization, 
\citet{Daigne} have
 developed models that include an early burst 
 of massive stars with several possible mass ranges, capable of reionizing the intergalactic medium, while satisfying observational constraints on 
cosmic chemical evolution in pre-galactic structures and in
the intergalactic medium.  In particular, 
\citet{Daigne} have demonstrated that 
the presence of massive stars ($M\sim$ 40-100 M$_\odot$)
is required  at high redshift ($z\gsim 15-20$). 
This early population of stars  (pop III)
is able to reionize the intergalactic  medium 
and generate a  prompt initial enrichment (PIE) in metals.
It is likely that particles will be accelerated
within the same process. 

Gamma-ray emission, as well as 
 cosmic rays,  may also come from active 
\citep{stecker, mukherjee} and 
normal \citep{pavlidou} galaxies
  \citep[see also][]{lemoine}. 
Depending on the strengh of 
the  magnetic fields in those structures,
 cosmic rays will be confined or will 
 propagate into the intergalactic medium 
\citep[e.g.][]{Berezinsky,Zweibel}.
In addition, recent numerical simulations
have shown that the formation of large scale structures
leads to accretion shocks in the baryonic gas, and thus to 
particle acceleration  directly in
the intergalactic medium \citep{kang,Miniati,keshet,ryu}.
Finally, at ultrahigh energies, more  exotic 
sources of cosmic rays have also been
studied  \citep{bhattacharjee,sigl}.
Clearly, there are several viable mechanisms for the production of
CCRs and just as clearly, there is a great deal of 
uncertainty surrounding their production.

In this paper,
CCRs are assumed to be 
produced in a single burst  correlated to a very early  generation of pop III stars as
discussed in \citet{Daigne}
at a given redshift \zs.
Note that very little  is known 
about the cosmic ray injection  spectra at these energies.
Here, our formalism is directly derived from 
the work of  \citet{Montmerlea}, hereafter M77. 
We briefly summarize this formalism 
and note explicitly our differences with this model.
A power-law distribution in particle
energy is adopted for the CR injection spectrum,
\begin{eqnarray}
\phi_\alpha(E)&=&\F\,12.5\,K_{\alpha p}(E+E_0)\{E(E+2\,E_0)\}^{-(\gamma+1)/2}\nonumber\\
 & &  {\rm cm}^{-2}{\rm s}^{-1}{\rm (GeV\ per\ nucleon)}^{-1}\,,
\EQN{phialpha}
\end{eqnarray}
which is the form expected from standard shock acceleration 
theory \citep{Blandford}. 
 $\F$ is a normalization factor which is fixed by the value of the
 injection spectral index, chosen to be  $\gamma=3$ \citep{Suzuki},
 and by $z_s$.
 It will ultimately be constrained 
by the observed abundance of \6Li\ in the MPHS (see \Sec{result}).
$E$ is the kinetic energy per nucleon, $E_0=939$ MeV  is the 
nucleon rest mass energy and $K_{\alpha p}=0.08$ is 
the abundance by number of $^4$He/H.
Lithium production is sensitive to $\alpha's$ with energy
$E \approx 10$ MeV/n.

\subsection{Transport function in an expanding universe}

The initial burst of cosmological cosmic rays
evolves in the framework of an expanding universe
with a cosmological constant.

If $N_i(E,z)$ is the comoving number density per (GeV/n) of a given
species at a given time or redshift, and energy, we define $N_{i,{\rm H}}(E,z)\, \equiv
\, N(E,z)/n_{\rm H}(z)$, the abundance by number with respect
to the ambient gaseous hydrogen (in units  of (Gev/n)$^{-1}$).
The evolution  of $N_{i,{\rm H}}$  is defined through the transport
function 
\begin{equation}
\frac{\partial N_{i,{\rm H}}}{\partial t} + \frac{\partial}
{\partial E}(bN_{i,{\rm H}}) + \frac{N_{i,{\rm H}}}{T_{\rm D}} = Q_{i,{\rm H}}\,.
\end{equation}
 $Q$ is a source function which accounts for 
different sources of particle production while 
${T_{\rm D}}$ is the lifetime against destruction.
$b$ describes the  energy losses due to expansion or ionization
processes ((Gev/n)\,s$^{-1}$).
The energy and time dependencies can be separated 
 as $b(E,z)=-B(E)f(z)$.
We can distinguish two cases depending on 
whether losses are dominated by expansion or by ionization. 
The general form for the redshift dependence, when 
expansion dominates
is $f_{\rm E}(z)=(1+z)^{-1}|\d{z}/\d{t}|\,H_0^{-1}$
\citep[e.g.][]{Wick}. Other contributions to 
$B$ or $f$, do not depend on the assumed cosmology
and are given explicitly in M77.

Two important quantities, $\zstar(E,E',z)$ and 
$\Eps(E,z)$ are used in this formalism. Given 
a particle ($\alpha$ or lithium) with an energy $E$ 
at a redshift $z$, $\zstar(E,E',z)$
corresponds to the redshift at which this particle
had an energy $E'$. $\Eps(E,z)$ is the initial energy required 
if this particle was produced  at the redshift of the burst,
$z_s$. In particular, $\zstar(E,\Eps,z)=z_s$.
The equation that defines $\zstar$ (Eq. A5, M77) is
$\partial{\zstar}/\partial{E} =  - \left[B(E)f(z)\,|\d{z}/\d{t}|\right]^{-1}\, \left(\partial{\zstar}/\partial{z}\right)$.
M77 gives analytical solutions for \zstar\ when \Ol\,=\,0.
When $\Ol\,\neq\,0$, it cannot be solved analytically 
when ionization dominates. Integration of this
equation 
shows that $\zstar(E,\Eps,z)$ is the solution of 
\begin{equation}
\int_z^{z^\star}\,\d{z''}\,f(z'')\,\left(|\d{z}/\d{t}|\right)_{z''}\ =\ \int_E^{{E_s^\prime}}\,\frac{\d{E''}}{B(E'')}
\EQN{zstar}
\end{equation}
 We solve this relation for \zstar\ numerically whenever
 analytical solutions
are not available.

\vskip 2cm

\subsection{The CCR flux and the lithium abundance}

The evolution of the CCR $\alpha$ energy  
spectrum is derived, using Eq. A8 of M77 and the single burst
properties, as 
\begin{equation}
\Phi_{\alpha, {\rm H}}(E,z) = {\phi_\alpha(E)\over \ n_{\rm H}^0}\frac{\beta}{\beta'}\frac{\phi_\alpha(\Eps)}{\phi_\alpha(E)}\left|\frac{\d{z}}{\d{t}}\right|_{z_s}\frac{\exp{(-\xi)}}{|b(E,z_s)|}\,\frac{1}{\left|\partial\zstar/\partial E'\right|_{{E_s^\prime}}}
\EQN{phialphaevol}
\end{equation}
where $\Phi_{\alpha, {\rm H}}(E,z)\equiv \Phi_\alpha(E,z)/n_{\rm H}(z)$
 is the flux of $\alpha$'s per comoving volume
\begin{equation}
\Phi_{\alpha, {\rm H}}(E,z) = \beta\,N_{\alpha,{\rm H}}(E,z)
\EQN{phialphadef}
\end{equation}
and  $\beta$ ($\beta'$)
is the velocity corresponding to energy
$E$ ($\Eps$); $\xi$ accounts for the destruction term (Eq. A9,
 M77).

The abundance by number of lithium
($l$\,=\,\6Li or \7Li) of energy $E$, 
produced at a given redshift $z$,
is computed from 
\begin{eqnarray}
{\partial N_{l,\, {\rm H}}(E,z) \over \partial t} & = & \int \sigma_{\alpha\alpha\rightarrow l}(E,E')n_{\rm He}(z)\Phi_{\alpha\, ,{\rm H}}(E',z)\,\d{E'} \nonumber\\
                      & = & \sigma_l(E) K_{\alpha p} \Phi_{\alpha}(4\,E,z)\ \ [\mbox{(Gev/n)$^{-1}$\,s$^{-1}$}]\,,
\end{eqnarray}
where $\sigma_{\alpha\alpha\rightarrow l}(E,E')=\sigma_l(E)\delta(E-E'/4)$. The cross sections used have been discussed in \Sec{observation}.
Note that 
this equation does not take into account the destruction of lithium 
in the intergalactic medium. 
We show below  that this is a reasonable
approximation.

Furthermore, we want to compute the abundance of lithium in the gas that 
is present at the redshift of the formation of the Galaxy (see below).
We assume that all the lithium produced will be thermalized in the protogalaxy 
before stars form. Thus, the quantity that should be compared
to the data is, 
\begin{equation}
 [l/{\rm H]}(z) = [l/{\rm H}]_{\rm BBN}+\int_z^{z^\star} \int {\partial N_{l,\, {\rm H}}(E,z')\over \partial t }\, \d{E}\, |\d{t}/\d{z'}|\, \d{z'}\,.
\EQN{lh}
\end{equation}
where $[l/{\rm H}]_{\rm BBN}$ is the primordial abundance 
predicted by  BBN.

The redshift evolution of \6Li/H
and \7Li/H  are 
the main results that will be 
compared to observations, and
used to constrain the CCR 
{\em proton} energy density,
\begin{equation}
{\cal E}_p(z) = \int_{E_{\rm cut}}\left[\Phi_\alpha(E_p,z)/K_{\alpha p}\right]\,E_p\, /\, \beta\ \d{E_p}.
\EQN{enerdens}
\end{equation}
where $E_{\rm cut}=10$ MeV 
 corresponds to the $\alpha$ energy cut-off (MeV/n)
 for the $\alpha+\alpha \rightarrow$ Li reaction.

Finally, we comment on the subsequent destruction of Lithium.
The differential rate of destruction (by protons) is equal to 
$\sigma_{\rm D}(E)\, n_{\rm H}(z)\, N_{l,\, {\rm H}}(E,z) \beta(E)$ and is proportional
to the Lithium abundance.
 The cross section $\sigma_{\rm D}$, decreases rapidly with 
energy below 10 MeV/n (see Fig. 2 of M77).
 Assuming  a constant energy of 10 MeV/n 
we can derive an upper limit
to the destruction process. 
We find  that taking  destruction 
into account increases the final proton energy density
only up to  7\% for \zs\ = 100 and has virtually no effect for $\zs \lsim 50$.

\subsection{Updated quantities}

Since 1977,  
the cosmological parameters, the \6Li\ and \7Li\
abundances predicted by BBN and observed in  MPHS
have changed considerably. They have been updated here.
Unless otherwise noted, we use the standard $\Lambda$CDM 
cosmology \citep{Spergel}
for which $H_0=71$ km\,\unit{s}{-1}\,\unit{Mpc}{-1},
 $\Om=0.27$, $\Ol=0.73$ with
$\Ob$h$^2$ = 0.0224.
The Hubble constant is defined as 
  $H(z) = H_0\,\left(\Om\,(1+z)^3+(1-\Omega_m - \Omega_\Lambda)(1+z)^2
  +\Ol\right)^{0.5}$ valid in a post-radiation dominated universe and we assume
  an equation of state parameter for the dark energy, $w = -1$. 
Finally, $\d{z}/\d{t}\, =\, -\, (1+z)\, H(z)$. 

We use the recently calculated BBN abundances of 
 lithium by \citet{Coc} and the
observed \6Li\  and \li7 abundances in MPHS.
{}From \Fig{obs}, we define the MPHS abundance
for \6Li\ as 
the value corresponding to the plateau, [$^6$Li]$_p$ = $\log($\li6/H$)+ 12 = 0.8$.

\section{Constraints on CCR production}
\label{s:result}

The process described above occurs in the IGM, and modifies
 the abundance pattern of the medium that will later form
 the Galaxy. Observations of MPHS trace the evolution 
of the gas in the halo of the Galaxy at an early epoch at
low metallicity. 
The observed abundance 
for the lowest metallicity is then assumed to be 
  a pregalactic 
abundance, i.e. the predicted abundance 
 at the redshift of the formation of the Galaxy.
This may be justified by the presence of the \6Li\
plateau. 
We will assume that the peak of the formation of 
the structures occurs at $\zgal\simeq 3$ \citep[e.g.][]{font,juneau,Hop}.
Therefore, the abundance observed for the lowest 
metallicity stars must correspond to the 
abundance in the intergalactic medium at 
$z=\zgal$. 
We next define the procedure used to constrain
the CCR burst parameters from the lithium observations.

\subsection{Procedure}
\label{s:procedure}

We begin by working within the context of the standard framework described in 
\Sec{formalism},  that is a $\Lambda$CDM + WMAP cosmology.
The shape of the CCR spectrum is
given by \Eq{phialpha}, with
$\gamma=3$.
Then, the evolution of the lithium abundance with redshift 
in our model is uniquely specified by the normalization
constant \F\ and the redshift of the CCR burst, \zs. 

Our CCR spectrum is constrained by $(i)$ the Spite plateau \citep{spites}
for \7Li\ and $(ii)$ the hint for a \6Li\ plateau \citep{Asplund2}.
The Spite plateau should correspond to the 
 primordial value of the \li7 abundance. 
However, the observed \7Li\ abundance is a factor of 2-3 lower
than the calculated one (based on the WMAP baryon density). 
As a result of this discrepancy, the \li7 plateau acts as a strong constraint in our model, 
since it forbids us to produce a non-negligible amount of \7Li.
This constraint is weakened if the observational value of 
\li7 were higher \citep[see e.g.][]{mr}
as the GCR component of 
\li7 would become more difficult to observe as the ratio of
GCR to BBN produced  \li7 is diminished  \citep{fieldsal}.
Nevertheless,  our model
must produce a small quantity of this isotope
compared to the BBN abundance.
The abundance of   \6Li\ 
observed at very low metallicity  is assumed to trace
the abundance in the \igm\ {\em before} the formation of the Galaxy.
As mentioned above, our model must be able to 
 reproduce this abundance
at $z=\zgal$. 

The constraints on the \6Li\ abundance from
 the calculated 
BBN abundance at $z\sim\infty$ and from  
its observed `pregalactic' value at $z=\zgal$
specify a unique
 amplitude for the CCR energy spectrum,   \F,
for a given \zs. Therefore, the normalization constant,
\F, is determined by the choice of parameters 
($\Ol,\,\Om,\, \gamma,\, \zs$), 
under the constraint given by the 
initial and final \6Li\ abundances. 
The initial \7Li\ abundance is fixed 
by the BBN. Then, for each
set of parameters, we check  that the
model does not produce too large of an additional
pregalactic component of \7Li.

\subsection{The  \6Li\ plateau and the evolution of \6Li\ versus redshift}

\begin{figure}[h]
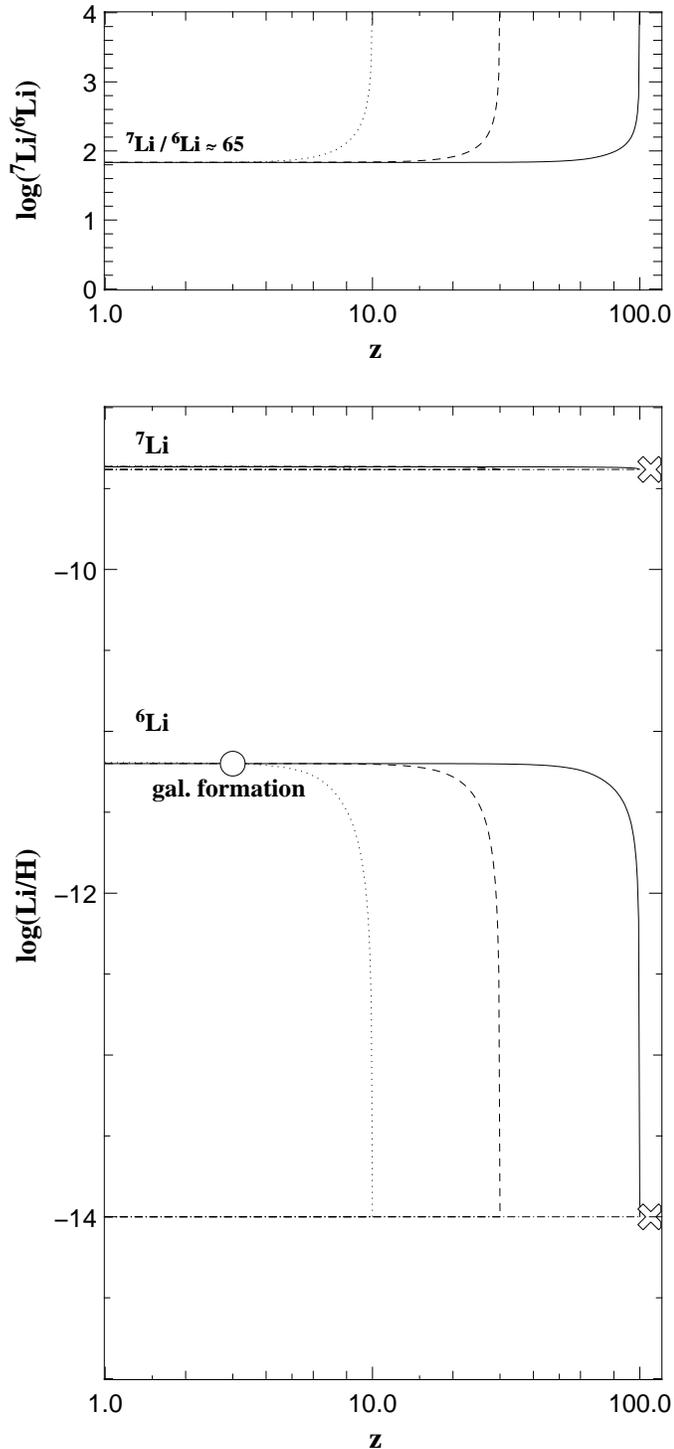
 
\unitlength=1cm
\begin{picture}(12,19.4)
\put(0,0){\psfig{width=\linewidth,figure=fig2.ps}}
\put(0,14.5){\psfig{width=\linewidth,figure=fig3.ps}}
\end{picture}
\caption{Redshift evolution of the \6Li\ and \7Li\
abundances in the intergalactic medium
(lower panel),
and of the ratio of \7Li/\6Li abundance (upper panel). 
The redshift of the initial CCR burst $z_s$
is chosen to be 10, 30 and 100 and is represented by the
dotted, dashed and solid lines respectively. 
The  shape of the CCR energy spectrum is fixed by
$\gamma=3$ and 
a $\Lambda$CDM cosmology \protect\citep{Spergel} is assumed.
The initial abundances of the lithium isotopes are
 fixed according to BBN calculations (crosses and horizontal 
dot-dashed lines)
while the abundance of \6Li\ 
is chosen to be $10^{-11.2}$ 
 at the redshift of the formation of the Galaxy,
$\zgal=3$ (circle). These choices fix the amplitude of the 
CCR flux (see \Sec{procedure}).
We find that the primordial abundance of \7Li\ 
 is increased by less than 10\%\ from \zs\ to \zgal.
}
\label{f:result}
\end{figure}

\parn The evolution of the abundance of \6Li\ 
with redshift is shown in \Fig{result} for three values
of $z_s=$10, 30 and 100.  
In each case a \6Li\ plateau is produced.
By construction, the abundance of \6Li\ at $z=\zgal$ is 
fixed. In addition, the rate of production of \li6 decreases rapidly
soon after the initial burst.
This was noted in M77
and corresponds to the dilution of the CCR flux with the expansion of the Universe.
Unless the burst occurs just prior to the formation of the Galaxy
(\Sec{discussion}), the \6Li\ abundance is almost constant for 
$z\lsim \zgal$.

Since the production rates of \6Li\ and
\7Li\ are similar, the additional production of \7Li\
due to CCRs ($\simeq\, 10^{-11}$) 
is negligible compared to the BBN primordial
values ($\simeq$ a few $ \times 10^{-10}$).
Note that  the predicted abundance in the Spite plateau
is increased by only 6\%, 8\%\ or 10\%\ for \zs=100, 30 or 10 respectively. 
The ratio \7Li/\6Li\ follows the same trend as 
\6Li\ (upper panel of \Fig{result}) and reaches
 a final value of about 60.

\subsection{Influence of the different parameters on the CCR amplitude}

We next investigate the influence of the parameters 
of the model ($\Ol$, $\Om$; 
$\gamma$, $z_s$) on the 
required amplitude of the CCR flux ($\F$) using
the same constraint on the  `pregalactic' value
[\li6]$_p = 0.8$  at $z=\zgal$. 
 Results are given in \Tab{result}.
Note that, as
mentioned above, the evolution of the lithium
production is
 dominated by the dilution of the CCR flux
that roughly follows a $(1+z)^3$ law. Thus,
the  cosmological parameters
 and the shape of the energy spectrum have very
little influence on the shape of the curves in \Fig{result}.
 Only the amplitude of the 
initial CCR flux varies.

 \begin{table}
 \caption{CCR amplitude versus input parameters}
 \begin{center}
 \begin{tabular}{c c c | c c }
 $z_s$ & ($\Omega_\Lambda$, $\Omega_m$) & $\gamma$ & $\log(\F)$ & $\log({\cal E}_p(z=\zs))$\\
100 & (0.7,0.3) & 3.0 & -4.6 & -10.2 \\
 & & 2.0 & -4.1 & -9.4  \\
 & (0.0,1.0) & 3.0 & -4.6 & -10.2 \\
 & & 2.0 & -4.0 & -9.3  \\
30  & (0.7,0.3) & 3.0 & -4.4 & -11.6 \\
 & & 2.0 & -3.7 & -10.6 \\
 & (0.0,1.0) & 3.0 & -4.1 & -11.4 \\
 & & 2.0 & -3.4 & -10.3 \\
10  & (0.7,0.3) & 3.0 & -3.5 & -12.2 \\
 & & 2.0 & -2.7 & -10.9 \\
 & (0.0,1.0) & 3.0 & -3.0 & -11.8 \\
 & & 2.0 & -2.3 & -10.5 \\
 \end{tabular}
 \end{center}
 The 
 redshift of the CCR burst (\zs), the cosmology, and the shape of the energy
 spectrum ($\gamma$) are varied. The amplitude of the 
 CCR spectrum (\F) is determined in order to reproduce the
  observed `pregalactic' ($z=\zgal$) abundance 
 [$^6$Li]$ _p = 0.8$  (\Fig{result}).
${\cal E}_p$ is the total initial energy density of protons in the CCRs 
(\Eq{enerdens}).
 \label{t:result}
 \end{table}

As one can see from the Table, 
 our results are
very sensitive to 
the redshift of the burst, $z_s$.
At high energy, one can show that $\Phi_{\alpha, {\rm H}}(E,z) \propto \phi_\alpha(E) \left(\frac{1+z}{1+\zs}\right)^{\gamma-1}$. 
Thus, the required  energy density at $z=\zs$
 in the  CCR is roughly proportionnal to 
$(1+\zs)^{1.5}$.
The CCR normalization also depends on the shape of the energy spectrum, $\gamma$. 
A steeper spectrum (higher $\gamma$)
favors the low energy part of the spectrum,
where the lithium production peaks. Thus, for 
a fixed amplitude \F, the abundance of \6Li\
will be higher for $\gamma=3$ than for 
$\gamma=2$.
Conversely, for a fixed abundance, $\cal F$
must be lower for $\gamma=3$.
Finally, there is little dependence on the cosmological parameters,
especially for large values of $\zs$.

We have used the observed plateau of \li6
to set the amount of pregalactic \li6 production.
Then, assuming a given epoch for the formation of 
CCRs, the amplitude of the energy spectrum, $\cal F$,
and the energy density are fixed (\Tab{result}).
The overall range of the proton energy density, ${\cal E}_p$
at $z=\zs$ 
is $10^{-10.2}$ to $10^{-12.2}$, for WMAP concordance model when $\gamma =3$ 
or $10^{-9.3}$ to $10^{-12.2}$ more generally. 
Yet, those CRs  may also play a role in
 heating and ionizing the IGM at high redshift.
 In fact,  when $z_s = 10$, the energy density of 6.3 $\times 10^{-13}$ ergs/cm$^{-3}$
 is marginally consistent with the resulting temperature of the IGM today.
 At higher $z_s$, this constraint is far less important as the resulting  IGM temperature
 scales as  $ {\cal E}_p /(1 + z_s)^{4}$ and since ${\cal E}_p$ increases slower than $(1+z_s)^4$.
 It is interesting to note that CCRs were predicted to heat the IGM and thus 
avoid the problem of overcooling in  the IGM gas \citep{blanchard}.
Furthermore, \citet{nath} have put other constraints on the 
total luminosity of the CRs from the Gunn-Peterson optical
depth at $z=4.2$, the Compton $y$-parameter and metal
enrichment. Assuming the production of CCRs from galaxies
at $z=10$, they obtained an initial
 luminosity of about $1.6\times 10^{-27}$
$h$ ergs/\unit{cm}{3}/s. Alternatively, from the 
amount of metals ejected by SN, they place an upper limit on the
cosmic ray energy density of $10^{-14}$ ergs/\unit{cm}{3} at $z = 0$ and
solar metallicity.  
At $z = 10$, the metallicity of the local ISM corresponding to the 
site of the CCR production is about .01 solar.  Note that this is much larger
than the resulting IGM metallicity where \li6 production occurs.
At $z = 10$ and at a metallicity of 0.01 solar, their limit is effectively
relaxed to ${\cal E}_p < 10^{-12}$ erg/cm$^{3}$.

Thus, we see that the CCR production of \li6 is capable of 
explaining the large abundance of \li6
with negligible production of \li7 (both relative to the BBN value).
Sufficient \li6 production is achieved by adjusting
the flux of CCRs and depends primarily on the assumed
redshift of the initial burst, $z_s$. 
Subsequently, the \li6 abundance remains roughly constant
until additional \li6 is produced in the Galaxy through
CGRs as we discuss in the next section.

 \section{Discussion}

\subsection{Galactic evolution of \6Li}
\label{s:discussion}

The above model for the CCR production can be thought of as 
a form of prompt initial enrichment if our Galaxy is formed
hierarchically from previously evolved structures. 
CCRs produce \li6 and a small amount (about a few \% of the BBN value)
of \li7.  Subsequently, the abundances of these element isotopes
as well as all other element abundances are controlled by
galactic chemical evolution. At this point, we assume only that the
initial abundances in the gas in the Galaxy correspond to those in the IGM at $z = z_{\rm gal}$.

It is certain that \li6 will  be produced in the ISM
through GCR nucleosynthesis (described briefly below)
as this is the primary mechanism for the production of 
\be9 and \b10. These isotopes will not have been produced in
any significant quantities as the IGM was initially devoid of 
C, N, and O needed for  spallation  processes.  In contrast, 
the presence of primordial \he4  allows for the CCR production of Li.

Cosmic rays produced in the early Galaxy will invariably interact with the
existing ISM.  
In standard GCR nucleosynthesis \citep{rfh}
 LiBeB nuclei are produced by spallation when
protons and $\alpha$s in the cosmic rays
impinge on ISM C, N, or O.
LiBeB is also produced when CNO in the cosmic rays are 
spalled by ISM protons and $\alpha$s.
As such,  spallation requires heavy elements (`metals')
to be  present in either the cosmic rays or the ISM.
In addition, $\alpha + \alpha$ fusion reactions between 
cosmic rays and the ISM lead to the production of 
the lithium isotopes.  Indeed, these were precisely the
types of process considered above in our model of 
CCR nucleosynthesis.
Note that \li7 and \b11 also receive contributions from the 
$\nu-process$ \citep{nuproc,opsv,vcfo}, but these will not be important for
our present discussion.

As in the case of CCR nucleosynthesis,  $\alpha$ fusion in
GCR nucleosynthesis is a primary process in contrast to 
the production of Be and B in standard models.  
The spallation of ISM CNO is a secondary process so
 these abundances scale as the square of a metallicity tracer such as O
or Fe if [O/Fe] is constant at low values of [Fe/H]. 
Motivated by the observational fact that 
the log of the Be and B abundances appear to scale 
linearly with [Fe/H]
 it has been proposed
that the bulk of cosmic rays are not accelerated
in the general ISM, but rather in the 
metal-rich interiors of superbubbles \citep{clvf,PD}, specifically at low metallicity.
Because the superbubble composition is
enriched in metals, any cosmic rays which are
accelerated in superbubble interiors would
have a composition which is both metal-rich and time-independent.
The low-energy component of the hard energy spectra
associated with superbubbles results in the primary production of Be and B. 
We refer to this process as LEC.
In general both the LEC and standard GCR nucleosynthesis are
responsible for the observed LiBeB abundances.

In Fig \ref{f:obs}, we display the GCR production of \li6 in the absence of 
the prompt initial enrichment produced by CCR nucleosynthesis. 
Here, we have normalized the flux of galactic cosmic rays 
so as to correctly reproduce the solar value of Be/H.  The overall flux is 
the only parameter available in GCR nucleosynthesis, and as a
consequence the abundances of \b10 and \li6 are predictions of the model
(recall that the \b11 and \li7 receive an additional contribution from the $\nu$-process).
As expected the logarithmic slope of [Li] versus [Fe/H] is 1 \citep{Fields99,Vangioni99}.

\begin{figure}[!h]
\unitlength=1cm
\begin{picture}(12,9)
\centerline{\psfig{width=\linewidth,figure=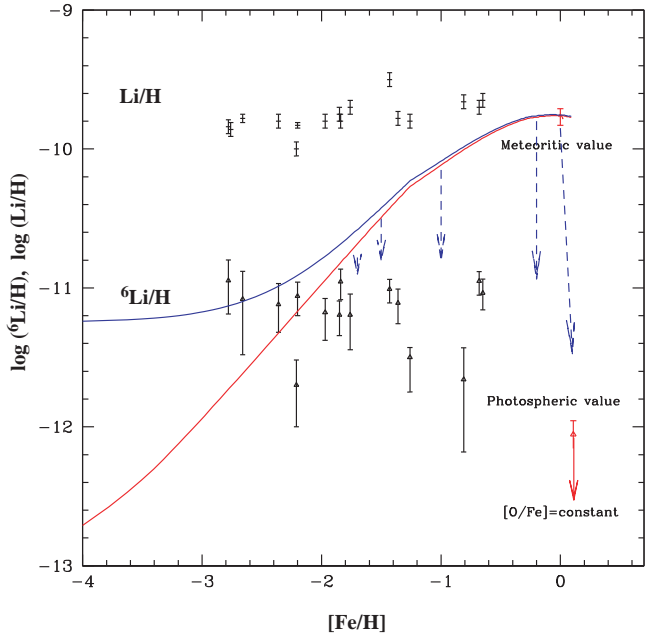}}
\end{picture}
\caption{
As in \Fig{obs}, the evolution of \6Li/H vs [Fe/H].
In this case the evolution of \6Li/H is modeled 
by GCR nucleosynthesis with the inclusion of  the LEC (lower line). 
Also shown is the case where a prompt initial enrichment is produced
 by CCRs in the intergalactic medium (upper line).
This model can explain the observed \li6 plateau at low metallicity.
The potential depletion of \li6  is indicated by the dashed arrows. 
}
\label{f:obs2}
\end{figure}

In \Fig{obs2}  we show the evolution of \6Li\ vs 
[Fe/H] when both CCR and LEC processes are included.  
 As one can see in Figure 3, 
without the initial enrichment of \li6 due to CCRs, the
evolution of \li6 resembles that of standard GCRs.
In this case, the \li6 abundance begins
at very low values and rises with a slope of unity until late 
times.  
This model alone can not explain the observational data.
At low [Fe/H] ($\la -2$), the observed \li6 abundance is too high to be accounted for by
standard CGR + LEC nucleosynthesis.

In addition, 
to explain the data at higher [Fe/H] ($\ga -2$), one must argue that
depletion has lowered the abundance of \li6.  This is perhaps reasonable as
the depth of the convection zone is increased at higher metallicity 
for a fixed surface temperature. We note that many 
of the stars observed only reveal upper limits to the \li6 abundance.
That is, in roughly 15 examples of stars with similar temperatures
and metallicities as those shown, no \li6 was detected. 
The lack of \li6 in some stars, coupled with the dispersion seen in the data
may also indicate that some depletion of \li6 has occurred in some of 
these stars. Indeed, 
the difference between the solar photospheric 
and meteoritic values corresponds to a destruction 
of \6Li\ of at least a factor of about 200. 
In this model, we would argue that the destruction of \6Li\ is negligible  
at [Fe/H] $\la -2$ where the calculation from galactic processes
cross the plateau. 

We also show in \Fig{obs2} the evolution of \li6 when the 
prompt enrichment due to CCRs is included.  
In this case, the data at low [Fe/H] is nicely modeled but 
depletion is still required to explain the data at higher metallicity.
At present, the evidence for the plateau hinges  on the abundances
in only a few stars at low metallicity.  However the two models shown in \Fig{obs2}
can be distinguished by future observations of \li6 at metallicities 
[Fe/H] $< -2.7$ which would establish the role of CCR nucleosynthesis
as a mechanism for the early production of \li6.
As mentionned above, 
a prompt initial enrichment (PIE)
in heavy elements is also expected 
 in the intergalactic medium, especially within
this Pop III stars scenario. However, the initial mass fraction
of iron may be of the order of X(Fe) = $10^{-7}$ \citep{Daigne}
and thus  do not  modify the curves in \Fig{obs} and \Fig{obs2},
 for [Fe/H]$\gsim\ -4$.
Thus, models with a \li6 plateau 
are not affected by the iron PIE generated by pop III stars.

The GCR nucleosynthesis models described above 
were based on the assumption that [O/Fe] was constant at low 
[Fe/H].  That is, we can use the iron abundance to trace
the evolution of the LiBeB elements. However some data
show that [O/Fe] increases with decreasing [Fe/H]
\citep{is,boes,ietal}.  As a consequence, the evolution of Be and B
may appear to be primary with respect to [Fe/H], but in fact is secondary
with respect to [O/H] \citep{fo}.  In reality, the data show that with the respect to 
[O/H], BeB have admixtures of primary and secondary components.
That is, the slope for log(BeB/H) vs. [O/H] is between 1 and 2 \citep{fovc,king}.

\begin{figure}[!h]
\unitlength=1cm
\begin{picture}(12,9)
\centerline{\psfig{width=\linewidth,figure=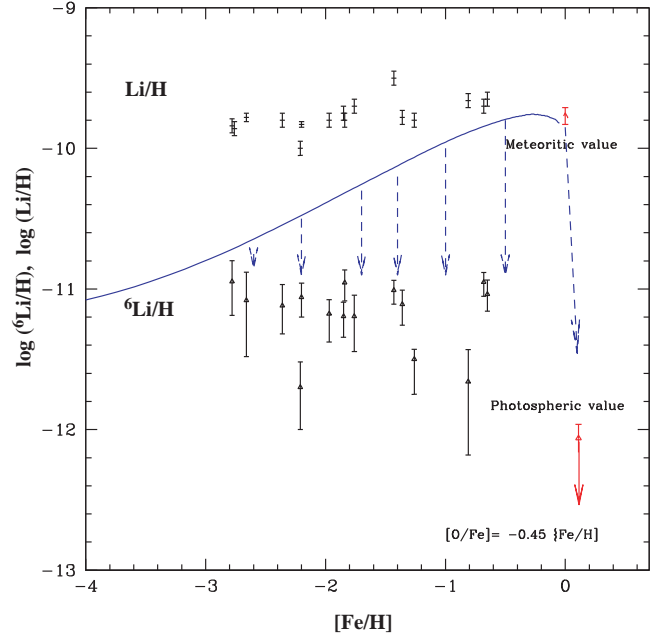}}
\end{picture}
\caption{The evolution of  \6Li/H vs [Fe/H]. 
Here, [O/Fe] is assumed to scale with [Fe/H].
We use
the maximum slope of the relation that
does not require  
a pre-galactic process (see text for details).
However, the amount of destruction required is much higher,
and must be present at metallicity as low as -3.
}
\label{f:OFe}
\end{figure}

In \Fig{OFe}, we show the resulting evolution of \li6 when the slope 
of [O/Fe] vs. [Fe/H] is taken to be -0.45 \citep[cf.][]{fovc,fieldsal}.
For a slope larger than -0.45, a prompt initial enrichment is not required since the 
galactic production of \li6 would already exceed the abundance observed in all stars.
However, within such a model, the destruction rate of 
\6Li\ must be non-negligible for metallicities as low as [Fe/H]$\simeq$ -3.
In \Fig{obs} and \Fig{obs2}, the slope was chosen to be 0 corresponding to constant [O/Fe].

\begin{figure}[!h]
\unitlength=1cm
\begin{picture}(12,9)
\centerline{\psfig{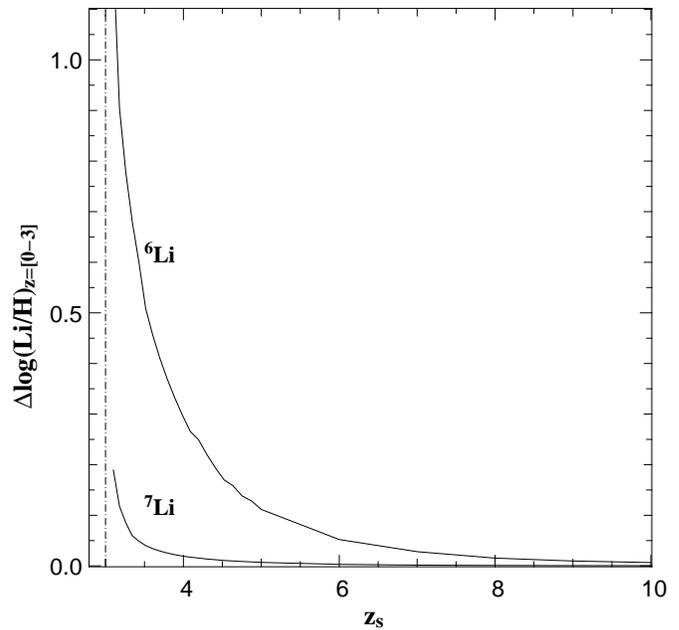}}
\end{picture}
\caption{Predicted dispersion in the  \6Li\  and \7Li\ abundances from $z=0$ to 3 as a function of
$\zs$. Note that by construction, $\zs$ must be greater than the redshift of formation of the Galaxy, $\zgal=3$ (vertical line).
}
\label{f:dispersion}
\end{figure}

Before concluding this part of the discussion, we note that the dispersion
in the data may be due to irregular production rather than depletion.
Due to the dilution of the CCR flux, the production of 
lithium saturates soon after the initial burst (\Fig{result}) at $z = z_s$. 
However, if this burst does not occur at a redshift
much larger than $\zgal$, the abundance of lithium 
can still increase. 
\Fig{dispersion} shows the expected variation of the \6Li\ and \7Li\ abundances from $z=0$ to 3.
If the IGM  can pollute the Galaxy \citep[e.g. through the merging of satellites, see e.g. ][]{navarro} 
at $z<\zgal$,
the abundance pattern in stars that form later may reflect this dispersion. 
For our choice of $z_{\rm gal} = 3$, the late production of Li  
may constrain the redshift of the CCR burst to be greater
than about 5-6.

\vskip 1cm

\subsection{Comparison with previous work}
\label{s:comparison}

\citet{Suzuki} consider  a 
model where cosmic rays generated by 
structure formation, during the process
of Galaxy formation, produce \6Li\  in the course of the evolution of 
the Galaxy. In their model, much of the \li6 is produced 
early and therefore they also
predict a \li6 plateau which extends down to at most  [Fe/H] $\approx -3$.
The characteristics of the plateau 
depend on the history of  structure formation.
If this process occurs early enough in the formation of the  Galaxy,
model I of \citet{Suzuki} is consistent
with the new observations (see their Fig. 1).
Once again, observations of \li6 at metallicities between -3 and -4
can distinguish between this model and the one we have presented here
for which the plateau is predicted to extend to much lower metallicities.
Furthermore, within the model of \citet{Suzuki}, the exact evolution of the \6Li\ abundance
 can be linked to the mean azimuthal rotation velocity
 of MPHSs. Consequently, they
predict larger dispersion among the observed \6Li\ abundances,
together with a  correlation between the \6Li\
abundance and 
the rotation velocity.
This could also help distinguish between galactic $\alpha+\alpha$ GCR  
production from the early $\alpha+\alpha$ CCR production
 considered in this paper.

The work by \cite{Fields}  uses a similar formalism 
to that described above, though redshift evolution is
not formally taken into account.
However, their main focus is on the lithium-gamma ray connection
in relation to the solar \6Li\ abundance.
Under this assumption, they claim that, 
if CCR interactions account for all of the \6Li\
production, it will also account for all of the observed 
extragalactic gamma-ray background (EGRB).
In this paper, we claim that CCRs must 
produce a  pregalactic \6Li\ abundance, that is about 
10 times smaller than the solar abundance
since most of the \6Li\ in stars at solar metallicity
is produced during galactic evolution (\Fig{obs}). 
Hence, we would argue that it should
produce only 10\% of the total EGRB (from their Eq. 13),
which is consistent with theoretical
predictions \citep[e.g.][]{Berezinsky,Colafrancesco,Miniati}.
\citet{PF} argue that the observation of Li in high-velocity clouds may
help establish the necessity of an early source of Li.

Finally, we note that \citet{Jedamzik,Jedamzik1,Jedamzik2} considers the very early production 
 of \6Li\ during BBN from decay \citep{Jedamzik,Jedamzik1} or annihilation \citep{Jedamzik1,Jedamzik2} of relic particles.
Naturally, this model will also predict the existence of an elevated plateau.

\section{Conclusion}
\label{s:conclusion}

The existence of the Spite plateau for \7Li\
indicates that low metallicity halo stars 
are representative of  the primordial BBN abundance,
although the discrepancy
with predictions based on WMAP results is still an issue 
\citep{Asplund,cfo4,Coc,Lambert, ryan04}.
On the other hand, the hint for a plateau in \6Li\ 
at very low metallicity, and at a higher abundance 
than predicted in standard BBN (by a factor of 1000), requires an additional
process that produces \6Li\  in a pregalactic 
phase.
The process studied in this paper involves
the interaction of $\alpha$ particles present in
early cosmological cosmic rays with 
primordial Helium present in the 
intergalactic medium.  We have shown that it is possible to 
produce sufficient quantities 
of \6Li, without  the additionnal over-production of \7Li.

The early production of \6Li\ 
will be present in the gas that forms the
Galaxy at $z=\zgal$ and
provides a simple explanation 
for the existence of 
the observed \6Li\ plateau in 
MPHS. The level of the \6Li\ plateau
may provide a strong constraint on the \zs-\F\ plane for
the initial burst of CCRs, and hence on
its total energy. 
However, the existence of this plateau
needs to  be confirmed 
 with additional observations of \6Li\ 
in stars with metallicities lower than -3. 
If the \li6 plateau persists down to lower metallicities, 
it could confirm the predictions of this model
and distinguish the physical processes occurring 
during Galaxy formation.

In a forthcoming paper, we will
go further than the single burst approximation. 
The CCR production could be related to the
formation and chemical evolution of 
pop III stars \citep{Daigne}.
The influence of this process in
the production of 
other elements, such as Be, B and D,
 will also be studied.

\acknowledgements
We are very grateful to Roger Cayrel for his always pertinent
 and fruitful comments. We thank E. Thi\'ebaut, and D. Munro for freely 
distributing his Yorick programming language (available at  {\em \tt
ftp://ftp-icf.llnl.gov:/pub/Yorick}), which we used to implement our
analysis.  
The work of ER was supported by a grant LAVOISIER from the French
foreign office.  The work of EVF has been supported by 
PICS 1076 CNRS France/USA. 
The work of K.A.O. was partially supported by DOE grant
DE-FG02-94ER-40823. 
\vskip 3.5cm

\end{document}